\documentclass[12pt]{article}
\usepackage{graphicx}
\def\g{{\cal G}^{++}}
\def\G{{\cal G}^{+++}}

\begin{document}
\thispagestyle{empty}
\setcounter{page}{0}

{\hfill{\tt hep-th/0511009}}

{\hfill{ULB-TH/05-22}}

\vspace{1cm}

\begin{center}{\bf \large Kac-Moody algebras in gravity and M-theories}\footnote{Talk presented at the XXVIII Spanish Relativity Meeting E.R.E. 2005, Oviedo, September 6
-10, 2005.}

\vspace{1cm}

Laurent Houart\footnote{Research  
Associate F.N.R.S.}

\vspace{.5cm}

 {\em Service de Physique Th\'eorique et Math\'ematique, Universit\'e Libre de Bruxelles,\\
  and\\ The International Solvay Institutes,\\
  Campus Plaine C.P. 231
  \\Boulevard du Triomphe, B-1050 Bruxelles, Belgium}

\vspace{1cm}

\end{center}

\centerline{\bf Abstract}

\noindent
The formulation of gravity and M-theories as very-extended Kac-Moody invariant theories is reviewed.
Exact solutions describing intersecting extremal brane configurations smeared in all directions but one
are presented. The intersection rules characterising these solutions are neatly encoded in the algebra. The existence of dualities for all $\G$ and their group theoretical-origin are discussed.

\newpage

A theory containing gravity suitably coupled to forms and dilatons may exhibit
upon dimensional reduction down to three dimensions a simple Lie group ${\cal G}$
symmetry non-linearly realised.
The scalars of the dimensionally reduced theory live in a coset
${\cal G}/{\cal H}$ where ${\cal G}$ is in its maximally non-compact form and
${\cal H}$ is the maximal compact subgroup of ${\cal G}$. A maximally oxidised theory is such a Lagrangian
theory defined in the highest possible space-time dimension $D$ namely a theory which is itself not
obtained by dimensional reduction. These maximally oxidised actions have been constructed for all   $\cal
G$ \cite{cjlp} and they include in particular pure gravity in $D$
dimensions and the low energy effective actions of the bosonic string and
of  M-theory.

\begin{figure}[h]
   \centering
   \includegraphics[width=10cm]{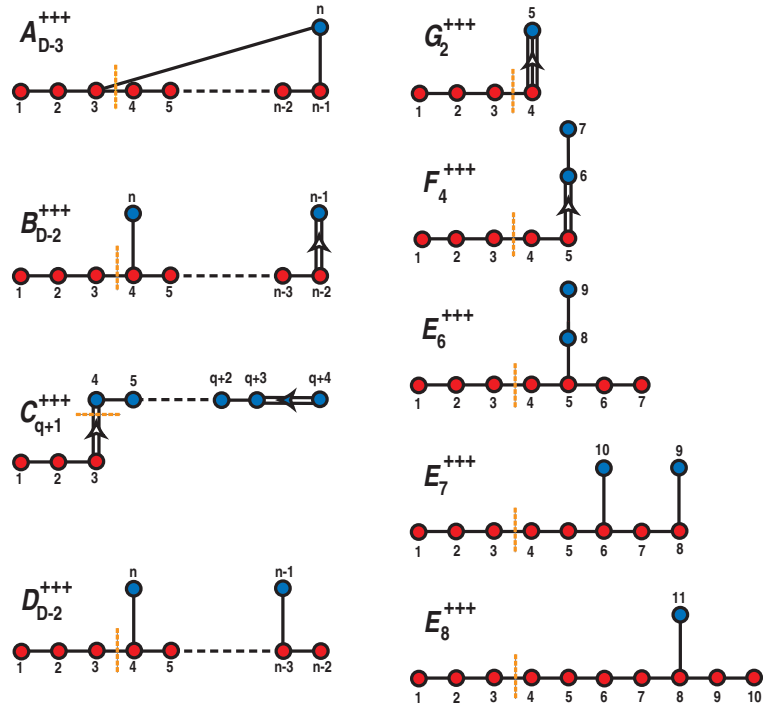}
 \caption { \small The
nodes labelled 1,2,3 define the Kac-Moody extensions of the  Lie
algebras. The horizontal line
starting at 1 defines the `gravity line', which is the
Dynkin diagram  of a
$A_{D-1}$ subalgebra.}
   \label{first}
\end{figure}

It has been conjectured that these theories, or some extensions of them,
possess the much larger very-extended Kac-Moody symmetry
$\G$. $\G$ algebras are defined by the Dynkin diagrams depicted
in Fig.1,  obtained from those of $\cal G$ by adding three
nodes~\cite{ogw}. One first adds the affine node, labelled 3 in the
figure, then a second node, 2,  connected to it by a single line and
defining the overextended ${\cal G}^{++}$ algebra\footnote{In the context of dimensional reduction,   the
appearance of
$E_8^{++}=E_{10}$ in one dimension has been first conjectured by B. Julia
\cite{julia84}.},   then  a third
one, 1, connected  by a single line to the overextended node. Such
$\G$ symmetries were first conjectured in the aforementioned
particular cases
\cite{west01,lw} and the extension to all
$\G$ was proposed in
\cite{ehtw}. In a different development, the study of the properties
of cosmological solutions in the vicinity of a space-like singularity,
known as cosmological billiards
\cite{damourhn00},  revealed an  overextended symmetry ${\cal G}^{++}$
for all maximally oxidised theories
\cite{damourh00, damourbhs02}.

The possible existence of this Kac-Moody symmetry $\G$ motivates
the construction of a Lagrangian formulation explicitly invariant
under $\G$  \cite{eh}. The action $S_{\G}$ is defined in a
reparametrisation invariant way on a world-line, a priori
unrelated to space-time, in terms of fields $\phi(\xi)$ living in
a coset $\G/K^{+++}$ where $\xi$ spans the world-line. A level
decomposition of $\G$ with respect to the subalgebra $A_{D-1}$ of
its gravity line (see Fig. 1) is performed where $D$ is
identified to the space-time dimension\footnote{ Level expansions
of very-extended algebras in terms of the subalgebra $A_{D-1}$
have been considered in \cite{west02, nifi, weke}.}. The
subalgebra  $K^{+++}$ is invariant under a `temporal' involution
which ensures that the action is $SO(1,D-1)$ invariant at each
level where the index $1$ of $A_{D-1}$ is identified to a time
coordinate.

Each $\G$ contains indeed a   subalgebra $GL(D)$ such that $SL(D) (=A_{D-1})
\subset GL(D) \subset
\G$. 
The generators of the $GL(D)$ subalgebra are taken to be
$K^a_{~b}\ (a,b=1,2,\ldots ,D)$   with commutation relations
\begin{equation}
\label{Kcom} [K^a_{~b},K^c_{~d}]   =\delta^c_b
K^a_{~d}-\delta^a_dK^c_{~b}\,  .
\end{equation}  The $K^a_{~b}$ along with  abelian generators $R_u \,( u=1 \dots q)$,
which are present  when the corresponding maximally oxidised action
$S_{\cal G}$ has $q$ dilatons\footnote{All the maximally oxidised  theories have at most one dilaton except
the $C_{q+1}$-series characterised by $q$ dilatons. In the rest of the paper we omit the $u$ index.}, are the level zero
generators. The step operators of level greater than zero are tensors
$R^{\quad c_1\dots c_r}_{ d_1\dots d_s}$ of the
$A_{D-1}$ subalgebra. Each  tensor forms an  irreducible representation of $A_{D-1}$ characterised
by some Dynkin labels.  In principle it is possible to determine the irreducible representations
present at each level \cite{nifi,weke}.
 The lowest levels contain antisymmetric tensor step  operators
$R^{a_1a_2 \dots a_r}$ associated to electric and magnetic
roots arising from the dimensional reduction of field strength forms in the corresponding maximally oxidised
theory. 
They satisfy the tensor and scaling relations
\begin{eqnarray}
\label{root} &&[K^a_{~b},R^{a_1\dots a_r}]   =\delta^{a_1}_b R^{aa_2
\dots a_r} +\dots +
\delta^{a_r}_b R^{a_1 \dots a_{r-1}a}\, ,\\
\label{root2} && [R, R^{a_1\dots a_r}
] =   -\frac{\varepsilon_A
a_A}{2}\,  R^{a_1\dots a_r}\, ,
\end{eqnarray} where $a_A$ is the dilaton coupling constant to the
 field strength form   and
$\varepsilon_A$ is $+1\, (-1)$ for an
electric (magnetic) root \cite{ehtw}.
The temporal involution $\Omega_1$ generalises the Chevalley
involution to allow identification of the index 1 to a time coordinate in
$SO(1,D-1)$. It is  defined by
\begin{equation}
\label{map} K^a_{~b}\stackrel{\Omega_1}{\mapsto}
-\epsilon_a\epsilon_b K^b_{~a}\quad R\stackrel{\Omega_1}{\mapsto} -R\quad
, \quad R^{\quad c_1\dots c_r}_{ d_1\dots d_s}
\stackrel{\Omega_1}{\mapsto}
-\epsilon_{c_1}\dots\epsilon_{c_r}\epsilon_{d_1}\dots\epsilon_{d_s}
  \bar R_{ c_1\dots c_r}^{\quad d_1\dots d_s}\, ,
\end{equation}
 with $\epsilon_a =-1$ if $a=1$ and
$\epsilon_a=+1$ otherwise. It leaves invariant a subalgebra $K^{+++}$ of
$\G$.
 The fields $\varphi(\xi)$ living in the coset space ${\G}/K^{+++}
$ parametrize the Borel group  built out of Cartan and positive
step operators in
$\G$. Its elements $\cal V$  are written as
\begin{equation}
\label{positive} {\cal V(\xi)}= \exp (\sum_{a\ge b}
h_b^{~a}(\xi)K^b_{~a} -
\phi(\xi) R) \exp (\sum
\frac{1}{r!s!} A^{\quad a_1\dots a_r}_{ b_1\dots b_s}(\xi) R_{
a_1\dots   a_r}^{\quad b_1\dots b_s} +\cdots)\, ,
\end{equation}
 where the first exponential contains only  level zero  operators and
the second one the positive step operators of levels strictly
greater than zero. Defining
\begin{equation}
\label{sym}
dv(\xi)= d{\cal V} {\cal V}^{-1}\quad d\tilde
v(\xi)=   -\Omega_1
\, dv(\xi)
\qquad\quad dv_{sym}=\frac{1}{2} (dv+d\tilde v)\, ,
\end{equation}
one obtains, in terms of the
$\xi$-dependent fields, an  action $S_{\cal \G}$ invariant under
global $\G$ transformations, defined on  the
coset ${\G}/K^{+++}$
\begin{equation}
\label{actionG} S_{\cal \G}=\int d\xi  \frac{1}{n(\xi)}\langle
(\frac{dv_{sym}(\xi)}{d\xi})^2\rangle\, ,
\end{equation} where
$n(\xi)$ is an arbitrary lapse function ensuring reparametrisation
invariance on the world-line and $< , >$ is the invariant bilinear form.

Writing
\begin{equation}
\label{full} S_{{\cal G}^{+++}} =S_{{\cal G}^{+++}}^{(0)}+\sum_A
S_{{\cal G}^{+++}}^{(A)}\, ,
\end{equation} where $S_{{\cal G}^{+++}}^{(0)}$ contains all level
zero contributions, one obtains
\begin{equation}
\label{fullzero} S_{{\cal G}^{+++}}^{(0)} =\frac{1}{2}\int d\xi
\frac{1}{n(\xi)}\left[\frac{1}{2}(g^{\mu\nu}g^{\sigma\tau}-
\frac{1}{2}g^{\mu\sigma}g^{\nu\tau})\frac{dg_{\mu\sigma}}{d\xi}
\frac{dg_{\nu\tau}}{d\xi}+
\frac{d\phi}{d\xi}\frac{d\phi}{d\xi}\right],
\end{equation}
\begin{equation}
\label{fulla} S_{{\cal G}^{+++}}^{(A)}=\frac{1}{2 r! s!}\int d\xi
\frac{ e^{- 2\lambda
\phi}}{n(\xi)}\left[
\frac{DA_{\mu_1\dots \mu_r}^{\quad \nu_1\dots
\nu_s}}{d\xi} g^{\mu_1{\mu}^\prime_1}...\,
g^{\mu_r{\mu}^\prime_r}g_{\nu_1{\nu}^\prime_1}...\,
g_{\nu_s{\nu}^\prime_s}
\frac{DA_{{\mu}^\prime_1\dots {\mu}^\prime_r}^{\quad
{\nu}^\prime_1\dots {\nu}^\prime_s}}{d\xi}\right].
\end{equation}  The $\xi$-dependent fields $g_{\mu\nu}$ are defined as
$g_{\mu\nu} =e_\mu^{~a}e_\nu^{~b}\eta_{ab}$ where $e_\mu^{~a}=(e^{-h(\xi)})_\mu^{~a}$. The appearance of the
Lorentz metric $\eta_{ab}$ with $\eta_{11}=-1$ is a consequence of the
temporal involution $\Omega_1$. The metric $g_{\mu\nu}$ allows a
switch from  the Lorentz  indices  $(a,b)$ of the fields appearing  in
Eq.(\ref{positive}) to
$GL(D)$ indices $(\mu,\nu)$.  $D/D\xi$ is a  covariant derivative generalising
$d/d\xi$ through  non-linear terms arising from  non-vanishing
commutators  between  positive  step operators and  $\lambda$ is the
generalisation of the scale parameter
$-\varepsilon_A a_A/2$ to all roots.

The  $\G$ -invariant actions $S_{{\cal
G}^{+++}}$ leads to  two distinct actions invariant under the overextended
Kac-Moody algebra $\g$. The first one $S_{\g_C}$ is
constructed from $S_{\G}$ by performing a consistent truncation.  The corresponding $\g$  algebra
is obtained from $\G$ by deleting the node labelled 1 from the
Dynkin diagram of $\G$ depicted in Fig. 1. The truncation is achieved by putting
to zero in the coset representative the field multiplying the Chevalley generator $H_1$ and all the fields
multiplying the positive step operators associated to roots whose decomposition in terms of simple roots contains the deleted root $\alpha_1$. This theory carries a
Euclidean signature and is the generalisation to all $\g$ of
the $E_8^{++}=E_{10}$ invariant action first proposed in reference
\cite{damourhn02}  in the context of M-theory and
cosmological billiards. The parameter $\xi$ is then identified
with the time coordinate and the action restricted to a defined
number of lowest levels is equal to the
corresponding maximally oxidised theory in which the fields depend
only  on this time coordinate. A second $\g$-invariant action
$S_{\g_B}$ is obtained from $S_{\G}$ by performing the same
consistent truncation {\it after} conjugation by the Weyl
reflection in the hyperplane perpendicular to  the simple root
 corresponding to the node 1 of figure 1. The
non-commutativity of the temporal involution with the Weyl
reflection \cite{keu1,keu2} implies that after Weyl reflection the index 2 of $A_{D-1}$ is now identified to the time coordinate. Consequently the second action $S_{\g_B}$  is
inequivalent to the first one \cite{ehh}.

 In $S_{\g_B}$,  $\xi$  is identified with a space-like direction and the action is characterised by
 a Lorentzian signature $(1,D-2)$. This theory admits exact solutions
which are identical to those of the corresponding maximally
oxidised theory describing intersecting extremal brane
configurations smeared in all directions but one
\cite{ehh,eh,eh2}.  
 For each of the $A=1 \dots \cal N$ branes
present in the intersecting brane configuration and characterised by
$\lambda_1 \dots \lambda_{q_A}$ longitudinal spacelike directions, one has
one non-zero field component corresponding to one positive step operator associated with one positive real
root $\alpha_A$.

\begin{equation} A_{2\lambda_1\dots \lambda_{q_A}}=
\epsilon_{2\lambda_1\dots \lambda_{q_A}}
[\frac{2(D-2)}{\Delta_A}]^{1/2}H_A^{-1}(\xi)
\qquad A=1 \dots {\cal N}\, ,
\label{aequ}
\end{equation} 

and
\begin{eqnarray}
\label{xx}
 &&p^a= \sum_{A=1}^{\cal N} p_A^a=\sum_{A=1}^{\cal N}
\frac{\eta_A^a}{\Delta_A}
\ln H_A(\xi)\qquad a=2,3,\dots,D  \\
\label{yy} &&\phi =\sum_{A=1}^{\cal N} \, \phi_A = \sum_{A=1}^{\cal
N}\frac{D-2}{\Delta_A}\varepsilon_A a_A
\ln H_A(\xi) \, ,
\end{eqnarray} where $p^a \equiv -h_a^{\quad a}$ and $h_a^{\, b}=0$ if $a \neq b$. 
Here $\eta_A^a=q_A+1$ or $-(D-3-q_A)$ depending on
whether  the direction $a$ is perpendicular or parallel to the
$q_A$-brane and
$\Delta_A= (q_A+1)(D-3-q_A)+\frac{1}{2}a_A^2(D-2)$.  The factor
$\varepsilon_A$ is $+1$ for an electric brane and $-1$ for  a magnetic one. Each
of the  branes in the configuration is thus described as electrically charged and
is characterised by one positive harmonic function  in
$\xi$-space, namely one has
\begin{equation}
\frac{d^2H_A(\xi)}{d\xi^2}=0 \qquad A=1 \dots {\cal N}\, .
\label{harmo}
\end{equation}
The consistent troncation yields for
the spatial  direction 1 the result
\begin{equation}
\label{dir1}
p^1= \sum_{A=1}^{\cal N}
\frac{q_A+1}{\Delta_A}\,
\ln H_A(\xi)\, ,
\end{equation}
identifying it to a direction transverse to all branes.
The Eqs. (\ref{xx}), (\ref{yy}) and (\ref{dir1})
are solutions provided the intersection rules
\cite{aeh}
\begin{equation}
\bar{q}+1=\frac{(q_A+1)(q_B+1)}{D-2}-\frac{1}{2}
\varepsilon_A a_A \varepsilon_B a_B \label{exorule}
\end{equation}
are satisfied. 

The intersection rules Eq.(\ref{exorule}) are neatly and elegantly encoded in the group
structure \cite{eh2}. They  can indeed be expressed
 as an orthogonality condition between the real positive roots of
$\g_B$ (and
$\G$) for all  branes present in the configuration
\cite{eh2} namely
\begin{equation}
\label{orthoroot}
\alpha_A \cdot \alpha_B=0 \qquad A \neq B=1 \dots \cal N
\end{equation}
When ${\cal G}$ is not simply laced there is an additional condition
in order to have a solution of $S_{\g_B}$ , namely for each pair $\alpha_A, \alpha_B$ with $A \neq B$ one
must have
\begin{equation}
\alpha_A + \alpha_B \neq {\rm root}.
\label{condau}
\end{equation}

The conditions Eqs. (\ref{orthoroot}) and (\ref{condau}) are in fact the input that
permits the derivation of the exact solutions by
allowing a reduction of $S_{\g_B}$ to  quadratic
terms. Furthermore there is a one-to-one correspondence between the exact solutions of $S_{\g_B}$ and
the  space-time intersecting extremal brane solutions of the corresponding maximally oxidised theory.
Indeed, the configurations satisfying Eq. (\ref{orthoroot}) and not Eq.(\ref{condau}) 
correspond in the maximally oxidised  theory to configurations which are not solutions
of the equations of motion because of the presence of Chern-Simons terms in the space-time action \cite{eh2}.

In the particular case of $\g =E_8^{++}$, it is well-known that the Weyl reflection generated by the root
$\alpha_{11}$ has an interpretation in terms of type IIA string duality. It correspond to a double T-duality in the direction 9 and 10 followed by an exchange of the two radii \cite{weylt, banks, ehtw}. Furthermore the change of signatures which occur as a consequence of the non-commutativity of the temporal involution and the Weyl reflections \cite{keu1,keu2} are in agreement with the exotic phases of M-theory discussed in \cite{hull1,hull2}.

The action of Weyl reflections generated by simple roots not belonging to the gravity line on the exact extremal brane solutions has been studied for {\it all} $\G$-theory \cite{eh}.
The existence of Weyl orbits of extremal brane
solutions similar to the U-duality orbits existing in M-theory has been discovered. This fact
strongly suggests a general group-theoretical origin of
`dualities' for all $\G$ -theories transcending string theories and
supersymmetry. Furthermore, exotic phases of all the M-theories (all the $\g_B$ theories) related by `duality' Weyl transformations to the conventional
phase\footnote{The other orbits have been discussed in \cite{keu3}.} characterised by a signature $(1,D-2)$ have been uncovered and classified \cite{bht}.

We exemplify the existence of U-duality-like Weyl orbits of extremal branes in the case 
 $\g_B= E_7^{++}$ whose corresponding maximally oxidised theory is  
 gravity  coupled to a 4- and a 2- form field strength  with one dilaton in 9 space-time dimensions \cite{cjlp}.
The Dynkin diagram of  
$E_7^{+++}$ is depicted in Fig.1, which exhibits the two simple electric
roots
$\alpha_{10}$ and $\alpha_{9}$ corresponding  respectively to the step operators
$R^{7\,8\,9}$ and $R^9$ which couple  to the electric potentials
$A_{7\,8\,9}$ and
$A_9$.

We take as input the electric extremal 2-brane ${\bf e}_{(8,9)}$ in the  
directions
$(8,9)$ associated with the 4-form field strength whose corresponding  
potential\footnote{We recall that in $\g_B$ the index 2 is identified to the time coordinate}  is
$A_{2\,8\,9}$ and submit it to the non trivial Weyl reflection $W_{10}$  
associated with the electric root
$\alpha_{10}$ of Fig.1. We display below, both for ${\bf e}_{(8,9)}$ and its  
transform, the vielbein components
$p^a \equiv - h_a^{\quad a}$ with $a=2 \dots 9$ and the the dilaton value $\phi$, of the brane solution  
Eqs.(\ref{xx}) and (\ref{yy}) as a nine-dimensional vector where the last
component is the   dilaton.  We also indicate the transform of the step
operator $R^{2\,8\,9}$ under   the Weyl transformation.   We obtain

\begin{eqnarray}
\label {e89} (-4,3,3,3,3,3, -4,-4;2\sqrt 7)\, \frac{\ln H(\xi)}{14}&  
{\bf e}_{(8,9)} &R^{2\,8\,9}\\ &\qquad\downarrow W_{10}\quad&\nonumber\\
\label{k17} (-7,0,0,0,0,7, 0,0;0)\, \frac{\ln H(\xi)}{14}& {\bf  
kk_e}_{\, (7)} &K^2_{~7}
\end{eqnarray} The 2-brane  transforms into a KK-wave in the direction 7 characterised by
a non-zero $K^2_{~7}$ \cite{eh}.  This  is  
reminiscent of a double T-duality  in   M-theory.

We now move the electric brane through Weyl reflections associated with  
roots of the gravity line to ${\bf e}_{(5 ,9)}$ and submit it to the Weyl  
reflection $W_{10}$. We now find that the brane ${\bf e}_{(5 ,9)}$ is 
invariant but   moving it to the position
${\bf e}_{(5 ,6)}$, we get
\begin{eqnarray}
\label {e56} (-4,3,3,-4,-4,3,3,3;2\sqrt 7)\, \frac{\ln H(\xi)}{14}&  
{\bf e}_{(5,6)} &R^{2\,5\,6}\\ &\qquad\downarrow W_{10}\quad&\nonumber\\
\label{m56789} (-1,6,6,-1,-1,-1,-1,-1;4\sqrt 7)\, \frac{\ln  
H(\xi)}{14}& {\bf m}_{(5,6,7,8,9)}  &R^{2\,5\,6\, 7\,8\,9}
\end{eqnarray} This is a magnetic 5-brane in the directions  
$(5,6,7,8,9)$ associated to the  2-form field strength~!   It is expressed
in terms   of its dual potential $A_{2\,5\, 6\,7\,8\,9}$. Submit instead
${\bf e}_{(5 ,9)}$ to   to the Weyl reflection $W_{9}$ associated with the
electric root $\alpha_9$ of Fig.1.   The 2-brane ${\bf e}_{(5   ,9)}$  is again 
invariant, but moving it to to the position
${\bf e}_{(5 ,6)}$, we now get
\begin{eqnarray} (-4,3,3,-4,-4,3,3,3;2\sqrt 7)\, \frac{\ln   H(\xi)}{14}&
{\bf e}_{(5,6)} &R^{2\,5\,6}\nonumber \\ &\qquad\downarrow  
W_{9}\quad&\nonumber\\
\label{m569} (-3, 4,4,-3,-3,4,4,-3;-2\sqrt 7)\, \frac{\ln   H(\xi)}{14}&
{\bf m}_{(5,6,9)}  &R^{2\,5\,6\, 9}
\end{eqnarray} This is a magnetic 3-brane in the directions $(5,6,9)$  
associated to the  4-form field strength, expressed in terms of its dual
potential  
$A_{2\,5\, 6\,9}$.

Finally, let us submit the magnetic 5-brane $ {\bf m}_{(5,6,7,8,9)}$  
obtained in Eq.(\ref{m56789}) to the Weyl reflection  $W_{9}$. One obtains
\begin{eqnarray} (-1, 6,6,-1,-1,-1,-1,-1;4\sqrt 7)\, \frac{\ln  
H(\xi)}{14}& {\bf m}_{(5,6,7,8,9)}  &R^{2\,5\,6\, 7\,8\,9}\nonumber \\
&\qquad\downarrow W_{9}\quad&\nonumber\\
\label{h2349} (0,7,7,0,0,0,0,-7;0)\, \frac{\ln H(\xi)}{14}&  {\bf  
kk_m}_{\, (1,3,4;9)} &R^{2\,5\,6\,7\,8\,9,\,9}
\end{eqnarray} Eq.(\ref{h2349}) describes, as in M-theory, a purely
gravitational     configuration, namely a KK-monopole  with transverse  
directions (1,3,4) and Taub-NUT direction 9 in terms of a dual gravity
tensor
$h_{2\,5\,6\,7\,8\,9,\,9}$ \cite{eh}.

 The  approach  based on Kac-Moody algebras constitutes certainly a very-exciting and innovative attempt to understand
 gravitational theories encompassing string theories  which could lead to a completely
 new formulation of gravitational interactions where the structure of space-time is hidden somewhere in these huge algebras \cite{damourhn02,ehh,dani} or even huger ones \cite{e12}.

\section*{Acknowledgments}

I would like to warmly thank my collaborator and friend Fran\c cois Englert with whom almost all the original results
presented here have been derived. I am also grateful to Sophie de Buyl, Marc Henneaux and Nassiba Tabti for enjoyable collaborations.

\end{document}